# IMPULSE DENOISING FROM HYPER-SPECTRAL IMAGES: A BLIND COMPRESSED SENSING APPROACH

Angshul Majumdar, Naushad Ansari, Hemant Aggarwal and Pravesh Biyani

{angshul, naushada, hemanta, pravesh}@iiitd.ac.in

Indraprastha Institute of Information Technology, Delhi

Abstract –

In this work we propose a technique to remove sparse impulse noise from hyperspectral images. Our algorithm accounts for the spatial redundancy and spectral correlation of such images. The proposed method is based on the recently introduced Blind Compressed Sensing (BCS) framework, i.e. it empirically learns the spatial and spectral sparsifying dictionaries while denoising the images. The BCS framework differs from existing CS techniques - which assume the sparsifying dictionaries to be data independent, and from prior dictionary learning studies which learn the dictionary in an offline training phase. Our proposed formulation have shown over 5 dB improvement in PSNR over other techniques.

## 1. INTRODUCTION

In this work, we address the problem of denoising hyper-spectral images when they are corrupted by impulse noise. Previously, most studies in hyper-spectral denoising only concentrated on removing Gaussian noise. However, hyperspectral images are also corrupted by impulse noise [1]; there is hardly any work on impulse denoising for hyper-spectral images.

There have been studies in removing impulse noise from gray-scale (single band) images. There are two approaches to address single band impulse denoising – traditional median filter based methods [2, 3] and modern optimization based techniques [4-6]. The optimization based techniques are more flexible in handling random valued impulse noise, while the median filtering variants are mostly suited for extreme valued salt-and-pepper noise.

Techniques developed for removing impulse noise from grayscale images can be applied on each spectral bands of the hyperspectral datacube, but such an approach would not be optimal. This is because, the hyperspectral images are spectrally correlated, and such piecemeal (band-by-band) denoising techniques do not account for the spectral correlation. Prior studies in Gaussian noise removal from hypercpetral images, showed that exploiting the spectral correlation indeed improves denoising results [7-9].

For denoising, the transform domain sparsity of the signal is usually exploited. Previously, data independent transforms (like wavelets) were used to sparsify the image; but in recent times it was observed that data dependent learned dictionaries yield better results for both Gaussian [10], impulse [5, 6] and even speckle [11] denoising.

In this work, we propose to exploit the spatio-spectral redundancy of the hyper-spectral datacube to reduce impulse noise. Such a problem has not been addressed before. However, instead of employing a fixed dictionary, we will learn the dictionary while denoising the hyperspectral datacube. Our approach is based on the Blind Compressed Sensing (BCS) approach [12]. Unlike prior studies [5, 6, 10, 11], we do not learn the dictionary in an offline stage and then use the learned dictionary for denoising; rather we learn the dictionary while denoising in an online fashion. This is the fundamental difference between BCS and prior dictionary learning techniques – BCS marries dictionary learning with signal estimation.

The rest of the paper will be organised into several sections. Relevant studies will be briefly reviewed in the next section. Our proposed approach will be described in section 3. The experimental results will be shown in section 4. The conclusions of this work and future direction of research will be discussed in section 5.

REVIEW OF LITERATURE

We are interested in the additive noise model; both Gaussian noise and impulse noise belong to this category. The noise model can be expressed as follows:

$$y = x + \eta \tag{1}$$

where x is the original image (to be estimated) corrupted to by noise η to yield a noisy image y.

Sprasity based techniques, assume that the image is sparse in a transform domain such as wavelets. Incorporating wavelet domain sparsity in (1) leads to:

$$y = W^T \alpha + \eta \tag{2}$$

where α is the sparse wavelet coefficients, W is the wavelet transform ($W^T$ is the inverse).

Gaussian denoising is the most well studied problem; assuming that the noise is zero mean with unit variance, denoising is achieved by solving for the sparse wavelet coefficients:

$$\min_{\alpha} \left\| y - W^T \alpha \right\|_2^2 + \lambda \left\| \alpha \right\|_1 \tag{3}$$

The $l_2$-norm data fidelity term arises owing to the nature of Gaussian noise; the $l_1$-norm regularization promotes a sparse solution. Wavelet sparsity is very well known topic in Gaussian denoising and does not require any reference for support.

There are also Total Variation (TV) [13] based denoising techniques that express denoising in the analysis form, i.e.

$$\min_x \|y - x\|_2^2 + \lambda TV(x) \qquad (4)$$

TV assumes that images are piecewise smooth; with finite number of jump discontinuities. This leads to a sparse representation in finite difference.

Gaussian noise is dense but usually small in magnitude - it corrupts all the pixel values but the magnitude of corruption is small. In such a case, an $l_2$-norm data fidelity is an ideal choice. Impulse noise is sparse but has larger amplitudes. An extreme case of impulse noise is the salt-and-pepper noise where the pixels are corrupted to either the maximum or minimum possible values; there can also be random valued impulse noise. In either case, as the number of corrupted pixels is small, an $l_1$-norm data fidelity is more appropriate [4, 14].

$$\min_x \|y - x\|_1 + \lambda TV(x) \qquad (5)$$

One can also formulate impulse denoising in the synthesis form, by exploiting wavelet domain sparsity instead of TV.

So far we have discussed techniques that exploit the sparsity of the image in a known basis (wavelet or finite difference). A seminal work [10] showed that, it is possible to improve upon these results by learning the sparsifying dictionary. In dictionary learning, the dictionary is first learnt from image patches, such that the learnt dictionary can represent the patches in a sparse fashion. The learning problem is expressed as,

$$\min_{D,Z} \|W - DZ\|_2^2 \text{ such that } Z \text{ is sparse} \qquad (6)$$

Here W is the training set of image patches, D is the learnt dictionary and Z is the sparse representation of the patches in dictionary D.

The learnt dictionary is later used for image denoising in the same manner as a wavelet dictionary, i.e. it is assumed that $x=Dz$. The denoising is expressed as:

$$\min_z \|y - Dz\|_2^2 + \lambda \|z\|_1 \qquad (7)$$

Recent studies in impulse denoising [5, 6] adopted the same technique. Since, the interest is in impulse noise, the data fidelity term is an $l_1$-norm instead of an $l_2$-norm.

$$\min_z \|y - Dz\|_1 + \lambda \|z\|_1 \tag{8}$$

So far, we have talked about denoising gray scale (single spectral band) images. In hyperspectral imaging, the image is acquired at multiple spectral bands. The images at different bands are correlated to one another. To reduce noise from such images, it is possible to apply the techniques developed for single band images to each of the spectral bands separately. But this does not yield the best possible results. Such techniques only exploit the spatial redundancies within each band; better results can be obtained by jointly exploiting the spatial and spectral correlations.

The noise model for hyperspectral images is as follows:

$$Y = X + N \tag{9}$$

Each column of X represents a clean hyperspectral image from a band; the columns of Y are the noisy versions of these images. The problem is to recover X given the noisy images Y.

It is well known that 2D wavelets lead to a sparse representation of images; this concept was extended to the hyperspectral datacube – in [15] it was shown that the datacube is sparse in 3D wavelets and hence denoising can be formulated as,

$$\min_\alpha \|vec(Y) - W_{3D}^T \alpha\|_2^2 + \lambda \|\alpha\|_1 \tag{10}$$

Here $\alpha = W_{3D} vec(X)$.

Similarly, there are other works that exploit the TV framework for denoising hyperspectral images using spatio-spectral correlations [8]. In some other studies [7], it is additionally assumed that the datacube X is of low-rank, this is because the columns of X are not linearly independent owing to spectral correlations.

PROPOSED DENOISING TECHNIQUE

Prior studies showed that exploiting spatio-spectral correlation of the hyperspectral datacube yields the better results than exploiting only spatial redundancies for Gaussian denoising. We follow the same principle while formulating our problem for impulse denoising. We assume, that the hyperspectral images X can be represented sparsely by two dictionaries $D_1$ (exploits spatial redundancy) and $D_2$ (exploits spectral correlation) in the following manner:

$$X = D_1 Z D_2 \tag{11}$$

Here, Z is the set of sparse coefficients.

In the standard sparse recovery framework, it is assumed that the dictionaries are given. Therefore, the recovery can be formulated as:

$$\min_{Z} \|Y - D_1 Z D_2\|_1 + \lambda \|Z\|_1 \tag{12}$$

As we studied before, $l_1$-norm data fidelity arises because impulse noise is sparse and affects only a few pixels. Here we have abused the notation slightly, the norms are defined on the vectorized versions of the matrices. But we do not use the 'vec' notation to keep the formulations uncluttered.

In the Blind Compressed Sensing (BCS) framework, the dictionary is learnt from the data; but unlike typical dictionary learning approaches, it is learnt during signal estimation and not in a separate offline training stage. Following the BCS framework, we apply a simple Tikhonov regularization on the dictionaries. This leads to the following problem:

$$\min_{Z, D_1, D_2} \|Y - D_1 Z D_2\|_1 + \lambda_1 \|Z\|_1 + \lambda_2 \|D_1\|_2^2 + \lambda_3 \|D_2\|_2^2 \tag{13}$$

This problem can be solved efficiently using the Split Bregman approach [16]. We introduce proxy variables – $P=D_1ZD_2$, $Q=Z$, $R=D_1$ and $S=D_2$. We add terms relaxing the equality constraints of each quantity and its proxy, and in order to enforce equality at convergence, we introduce Bregman variables $B_1$, $B_2$, $B_3$ and $B_4$. The new objective function turns out to be:

$$\min_{Z, D_1, D_2, P, Q, R, S} \|P\|_1 + \lambda_1 \|Q\|_1 + \lambda_2 \|R\|_2^2 + \lambda_3 \|S\|_2^2 + \mu \|P - (Y - D_1 Z D_2) - B_1\|_2^2 \\ + \mu_1 \|Q - Z - B_2\|_2^2 + \mu_2 \|R - D_1 - B_3\|_2^2 + \mu_3 \|S - D_2 - B_4\|_2^2 \tag{14}$$

The variable splitting allows us to express (14) as an alternating minimization of the following (easier) sub-problems:

$$P1: \min_{Z} \mu \|P - (Y - D_1 Z D_2) - B_1\|_2^2 + \mu_1 \|Q - Z - B_2\|_2^2 \tag{15a}$$

$$P2: \min_{D_1} \mu \|P - (Y - D_1 Z D_2) - B_1\|_2^2 + \mu_2 \|R - D_1 - B_3\|_2^2 \tag{15b}$$

$$P3: \min_{D_2} \mu \|P - (Y - D_1 Z D_2) - B_1\|_2^2 + \mu_3 \|S - D_2 - B_4\|_2^2 \tag{15c}$$

$$P4: \min_{P} \|P\|_1 + \mu \|P - (Y - D_1 Z D_2) - B_1\|_2^2 \tag{15d}$$

$$P5: \min_{Q} \lambda_1 \|Q\|_1 + \mu_1 \|Q - Z - B_2\|_2^2 \tag{15e}$$

$$P6: \min_{R} \lambda_2 \|R\|_2^2 + \mu_2 \|R - D_1 - B_3\|_2^2 \tag{15f}$$

$$P7: \min_{S} \lambda_2 \|S\|_2^2 + \mu_3 \|S - D_2 - B_4\|_2^2 \tag{15g}$$

Apart from P4 and P5, the rest are least squares minimization problems which can be solved efficiently using Conjugate Gradient techniques. The sub-problems P4 and P5 are $l_1$-norm regularized least squares problems which can be solved via iterative soft thresholding [17].

The final step of the Split Bregman technique is to update the relaxation variables:

$$B_1 \leftarrow P - Y + D_1 Z D_2 - B_1 \tag{16a}$$

$$B_2 \leftarrow Q - Z - B_2 \tag{16b}$$

$$B_3 \leftarrow R - D_1 - B_3 \tag{16c}$$

$$B_4 \leftarrow S - D_2 - B_4 \tag{16d}$$

There are two stopping criterions for the Split Bregman algorithm. Iterations continue till the objective function converges (to a local minima; the BCS formulation is non-convex thus there is no guarantee for global convergence); by convergence we mean that the difference between the objective functions between two successive iterations is very small ($10^{-4}$). The other stopping criterion is a limit on the maximum number of iterations. We have kept it to be 200.

EXPERIMENTAL RESULTS

Two hyperspectral datacubes were used for performing experiments. One is of Reno city, NV, USA available from [22]. This image is from High Resolution Imager (HRI) sensor having 2m spatial resolution and 5 nm band spacing covering spectral range of 395-2450 nm. The second dataset of of Washington DC mall available from [23]. This image is of Hyperspectral Digital Imagery Collection Experiment (HYDICE) sensor having 1m spatial resolution and 10-nm band spacing covering spectral range of 400-2500 nm. We used patches of size 64 × 64 × 64 from both the images for experiments. A portion of the pixels were corrupted by random valued noise. The task is to recover the underlying image from the noisy version.

We have experimented with several denoising algorithms. The $l_1$-KBCS stands for Kronecker Blind Compressed Sensing with $l_1$-norm data fidelity constraint - this is our proposed technique (13). The $l_1$-BCS stands for BCS with $l_1$-norm for data fidelity, i.e. it only sparsifies along the spatial direction but not along the spectral direction. We achieve this by putting $D_2$ = Identity and $\lambda_3$=0 in (13). MF stands for the standard median filtering technique applied to each of the bands separately. The $l_1$-TV stands applies the impulse denoising formulation proposed in [4]. The $l_1$-KCS formulation is the one proposed in (12) - it uses fixed dictionaries (wavelets) to sparsify along the spatial and spectral directions.

Our proposed method requires specifying several parameters. The parameters were tuned on a validation set. For tuning the parameters we employed a sub-optimal yet effective strategy based on the L-curve method. For the first parameter (say $\lambda_1$) we set the other parameters to zero and use the L-curve method to find it. To tune the second parameter $\lambda_2$, we fix $\lambda_1$ to the obtained value and the remaining parameters to zero; we again use the L-curve method to determine $\lambda_2$. The same strategy is repeated until all the parameters are determined. We varied the parameters on the log scale (100, 10, 1, 0.1 etc.). We obtained the following values: $\lambda_1=10^{-1}$, $\lambda_2=10^{-1}$ and $\lambda_3=10^4$.

Our algorithms requires specifying some hyper-parameters. The Bregman relaxation variables were randomly (uniform distribution between zero and one) initialized for both the datasets. The internal variables ($\mu$'s) were fixed by trial and error to yield the best results on the validation dataset. As before, we only varied the values on a log scale. The following values were used: $\mu=10$, $\mu_1=10$, $\mu_2=10^3$ and $\mu_3=10^3$.

Table 1. PSNR from Various Techniques

The results from Table 1 show that our proposed algorithm always yields very good results. When the noise is small, the $l_1$-BCS formulation (i.e. without exploiting spectral correlation) yields slightly better results than the $l_1$-KBCS algorithm. But when the noise increases, it cannot keep up; the results deteriorate drastically. Our proposed $l_1$-KBCS formulation that exploits sparsity along both the spatial and spectral directions yields the best results. Usually in image denoising problems, 0.5 - 1 dB improvement in SNR is considered significant; in this work our proposed method improves the results over existing techniques by more than 5 dB. It should be kept in mind, that MF and l1-TV are the only existing studies for impulse noise removal; the rest are proposed in this paper. Compared to the existing works, we get more than 10 dB improvements.

For visual assessment, the images are shown in Fig. 1. We show the Reno image from a randomly chosen spectral band with 30% corrupted pixels. We find that the best reconstruction - maximum noise removal with minimum blurring, is achieved by our proposed $l_1$-KBCS method. The $l_1$-KCS and $l_1$-BCS methods yields almost similar outcomes even though in terms of PSNR, the $l_1$-BCS technique seem to show worse results. The median filtering techniques cannot remove all the noise. The $l_1$-TV technique removes noise but results in blurring.

Fig. 1. Left to Right: Ground-truth, Noisy Image, MF, $l_1$-BCS, $l_1$-KBCS, $l_1$-TV and $l_1$-KCS

CONCLUSION

This work proposes a technique to reduce impulse noise from hyper-spectral images. The formulation is based on the fact that impulse noise is sparse and hence the data fidelity term should be an *$l_1$*-norm. It accounts for spatio-spectral correlation in the hyperspectral datacube by using separate dictionaries to sparsify along the spatial and spectral directions. But, rather than using a fixed dictionary, our technique empirically learns the dictionary from the data while denoising. Loosely speaking, our work is based on the Blind Compressed Sensing (BCS) framework [12].

We have compared our proposed method with other competing techniques, and have shown that our proposed method outperforms others by a considerable margin. The experiments were carried out on real datasets with varying degrees of noise.

In CS, the number of measurements should be smaller than the dimensionality of the signal to be recovered. Therefore strictly speaking, denoising is not a Compressed Sensing (CS) problem. However, our proposed formulation can be used to reconstruct hyperspectral images from compressive measurements in the presence of impulse noise. There have been some studies on compressive hyperspectral imaging in the presence of Gaussian noise [18, 19]. In the past, BCS and its variants have been successful in recovering dynamic Magnetic Resonance Imaging sequences from under-sampled measurements [20, 21]. We expect a similar success in compressive hyperspectral imaging.

The implementations for reproducing these results can be obtained from the authors on request. We plan to make these available in Matlab Central soon.

Table 1. PSNR from Various Techniques

| Name of Algorithm | 10% Noise | | 30% Noise | | 50% Noise | |
|---|---|---|---|---|---|---|
| | WDC mall | Reno | WDC mall | Reno | WDC mall | Reno |
| MF | 29.62 | 35.98 | 21.17 | 23.16 | 14.78 | 14.13 |
| $l_1$-BCS | **55.24** | **52.06** | 37.49 | 38.05 | 8.92 | 9.00 |
| $l_1$-KBCS | 53.82 | 48.30 | **48.64** | **48.25** | **39.93** | **38.45** |
| $l_1$-TV | 41.09 | 35.27 | 39.51 | 32.57 | 37.38 | 29.27 |
| $l_1$-KCS | 45.75 | 36.96 | 43.27 | 35.04 | 37.25 | 27.68 |

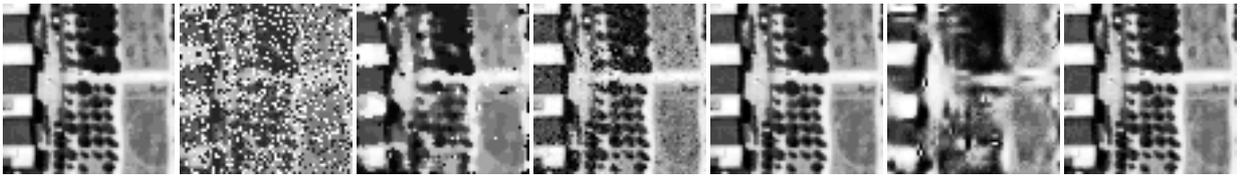

Fig. 1. Left to Right: Ground-truth, Noisy Image, MF, $l_1$-BCS, $l_1$-KBCS, $l_1$-TV and $l_1$-KCS